\documentclass[a4paper]{jpconf}

\usepackage{graphicx}
\usepackage{dcolumn}
\usepackage{bm}
\usepackage{epsfig}
\def\b{\begin{equation}}
\def\e{\end{equation}}

\def\hedm{H_{\rm EDM}}

\makeatletter
 \def\overbracketnew#1{\mathop{\vbox{\ialign{##\crcr\noalign{\kern6\p@}
   \downbracketfill\crcr\noalign{\kern6\p@\nointerlineskip}
   $\hfil\displaystyle{#1}\hfil$\crcr}}}\limits}
 \def\downbracketfill{$\m@th
   \kern6\p@ \makesm@sh{\llap{\vrule\@height.9\p@\@depth2.3\p@\@width.9\p@}}%
   \leaders\vrule\@height.9\p@\hfill\kern 14\p@
   \makesm@sh{\rlap{\kern-14\p@\vrule\@height.9\p@\@depth2.3\p@\@width1.\p@}}$}
 \def\overbracket#1{\mathop{\vbox{\ialign{##\crcr\noalign{\kern3\p@}
   \downbracketfill\crcr\noalign{\kern3\p@\nointerlineskip}
   $\hfil\displaystyle{#1}\hfil$\crcr}}}\limits}
 \def\downbracketfill{$\m@th
   \kern4\p@ \makesm@sh{\llap{\vrule\@height.7\p@\@depth2.3\p@\@width.7\p@}}%
   \leaders\vrule\@height.7\p@\hfill\kern 14\p@
   \makesm@sh{\rlap{\kern-14\p@\vrule\@height.7\p@\@depth2.3\p@\@width.7\p@}}$}
 \def\overbracketl#1{\mathop{\vbox{\ialign{##\crcr\noalign{\kern3\p@}
   \downbracketfilll\crcr\noalign{\kern3\p@\nointerlineskip}
   $\hfil\displaystyle{#1}\hfil$\crcr}}}\limits}
 \def\downbracketfilll{$\m@th
   \kern4\p@ \makesm@sh{\llap{\vrule\@height1.1\p@\@depth2.3\p@\@width1.1\p@}}%
   \leaders\vrule\@height1.1\p@\hfill \kern4\p@
   \makesm@sh{\rlap{}}$}
 \def\overbracketr#1{\mathop{\vbox{\ialign{##\crcr\noalign{\kern3\p@}
   \downbracketfillr\crcr\noalign{\kern3\p@\nointerlineskip}
   $\hfil\displaystyle{#1}\hfil$\crcr}}}\limits}
 \def\downbracketfillr{$\m@th
   \kern4\p@ \makesm@sh{\llap{}}%
   \kern-6\p@\leaders\vrule\@height.7\p@\hfill \kern14\p@
   \makesm@sh{\rlap{\kern-14\p@\vrule\@height.7\p@\@depth2.3\p@\@width.7\p@}}$}
 \def\overbrackettw#1{\mathop{\vbox{\ialign{##\crcr\noalign{\kern3\p@}
   \downbracketfilltw\crcr\noalign{\kern3\p@\nointerlineskip}
   $\hfil\displaystyle{#1}\hfil$\crcr}}}\limits}
 \def\downbracketfilltw{$\m@th
   \kern4\p@ \makesm@sh{\llap{\vrule\@height.7\p@\@depth2.3\p@\@width.7\p@}}%
   \leaders\vrule\@height.7\p@\hfill\kern 14\p@
   \makesm@sh{\rlap{\kern-14\p@\vrule\@height.7\p@\@depth7.8\p@\@width.7\p@}}$}
 \def\overbracketth#1{\mathop{\vbox{\ialign{##\crcr\noalign{\kern3\p@}
   \downbracketfillth\crcr\noalign{\kern3\p@\nointerlineskip}
   $\hfil\displaystyle{#1}\hfil$\crcr}}}\limits}
 \def\downbracketfillth{$\m@th
   \kern4\p@ \makesm@sh{\llap{\vrule\@height.7\p@\@depth2.3\p@\@width.7\p@}}%
   \leaders\vrule\@height.7\p@\hfill\kern 14\p@
  \makesm@sh{\rlap{\kern-14\p@\vrule\@height.7\p@\@depth14.0\p@\@width.7\p@}}$}
\makeatother

\def\ob{\overbracket}

\begin{document}

\title { A new formulation of the relativistic many-body theory of electric 
         dipole moments of closed shell atoms}

\author{K V P  Latha$^1$, Dilip Angom$^2$, Rajat K Chaudhuri$^1$ 
        ,B P Das$^1$ and Debashis Mukherjee $^3$}
\address{$^1 $ Indian Institute of Astrophysics,
Bangalore - 560 034, India}

\address{$^2 $ Physical Research Laboratory,
                   Navarangapura, Ahmedabad}
\address{$^3$ Indian Association of Cultivation of Science,
                           Kolkata, West Bengal}

\ead{latha@iiap.res.in}

\begin{abstract}
 
   The electric dipole moments of closed-shell atoms are sensitive to the 
parity and time-reversal violating phenomena in the nucleus. The nuclear 
Schiff moment is one such property, it arises from the parity and time reversal 
violating quark-quark interactions and the quark-chromo electric dipole 
moments. We calculate the electric dipole moment of atomic $^{199}{\rm Hg}$ 
arising from the nuclear Schiff moment using the relativistic coupled-cluster 
theory. This is the most accurate calculation of the quantity to date. Our 
calculations in combination with the experiment data provide important 
insights to the P and T violating coupling constants at the elementary particle 
level. In addition, a new limit on the tensor-pseudo tensor induced atomic 
EDM, calculated using the relativistic coupled-cluster theory is also 
presented.

\end{abstract}



\section{Introduction}

The origin of parity violation can be explained within the frame work of the 
Standard Model (SM) of particle physics through the weak interactions, but 
there is no clear understanding of the origin of time-reversal violation in 
nature. The presence of a non-zero electric dipole moment of a non-degenerate 
physical system is a direct signature of parity ($P$) and time-reversal ($T$)
symmetry violations \cite{commins,lee,landau}. In this paper we present a study of atomic 
EDMs of closed-shell atoms which arise mainly from the tensor-pseudo tensor (T-PT) electron-nuclear 
interactions and the nuclear Schiff moment (NSM). At the elementary
particle level, the origin of closed-shell atomic EDMs is attributed to the
$P$ and $T$ violating electron-quark interactions and quark-quark interactions
which are predicted by the lepto-quark models, SUSY, etc \cite{barr}. The 
limits on the T-PT coupling constant ($C_T$) has been obtained from the 
comparison of the most recent experimental result of $^{199}{\rm Hg}$ atomic EDM 
\cite{roma,jacobs}, $|d(^{199}{\rm Hg})|< 2.1\times 10^{-28} {\rm e cm }$
and the enhancement factor (ratio of atomic EDM to the coupling constant of 
the interaction in question) calculated in \cite{ann-mart}. The calculation uses
Coupled-perturbed Hartree-Fock (CPHF)\cite{peter,caves} theory and the calculated value of
the atomic EDM is $d(^{199}{\rm Hg}) = - 6.0 \times 10^{-22}  C_T \sigma_N{\rm e m}$
combining with experimental result, a limit on $C_T$ can be derived,
$$
   C_T < 3.26  \times 10^{-9} \sigma_N
$$
where $\sigma_N $ is the nuclear spin.
A non-zero value of $C_T$ would imply physics beyond the Standard Model. Improving
the accuracy of the calculations would lead to a more 
stringent bound on $C_T$. The CPHF theory accounts for two-particle two-hole type 
of electron correlations to all orders in perturbation. In addition to this, 
for accurate atomic calculations\cite{liu}, it is important to include other correlation 
effects which are absent in the CPHF theory. Coupled-cluster (CC) theory \cite{coester,bart}
is one of the many-body methods which is non-perturbative and incorporates all 
forms of correlation effects \cite{kaldor,derevianko,bart_rev}. CC theory with all levels of
excitations is equivalent to all order Many-body Perturbation Theory. It
has been successfully applied for calculation of a variety of atomic properties \cite{derevianko,
holger,kim}. It has also been applied to atomic Hg for calculating transition energies \cite{kaldor}.

  The NSM(denoted by operator $\vec S$) can originate from the nucleon-nucleon 
interactions or a nucleon EDM. At the elementary particle level it can
arise from the interaction between the quarks and the chromo EDM
of the quarks. The coupling constants associated with these 
interactions can be predicted by Multi-higgs, SUSY
\cite{barr,falk} etc. The dependence of the T-PT and NSM interactions on the 
nuclear spin makes closed-shell atoms, in particular, those having non-zero 
nuclear spin the best candidates to look for EDMs sensitive to the nuclear 
sector. For $^{199}{\rm Hg}$, the EDM induced by the NSM is calculated and 
parameterized in terms of the Schiff moment operator $\vec S$. The most recent 
calculation \cite{dzuba,flamrevw} gives
$$
  d_{\rm Hg} = - 2.8 \times 10^{-17} \left(\frac{S_{\rm Hg}}{\rm e fm^3}\right) 
  {\rm e cm}.
$$
According to the recent work \cite{flamging} the expression for the 
P and T - odd nuclear potential arising from the nuclear Schiff moment
\begin{equation}
  \Phi(\vec R) = -3  \frac{\vec S \cdot \vec R}{B}\rho(R) ,
\end{equation}
where $B=\int{R^4\rho(R) dR}$ and $R$ is the electron coordinate and $\rho(R)$ is
the nuclear density.
The interaction Hamiltonian of electrons with this potential is
\begin{equation}
  H_{\rm SM} = 3 {\rm e} \frac{\vec S \cdot \vec R}{B}\rho(R) 
\label{phism}
\end{equation}
where $e$ is the charge of the electron.
If $\rho(R)$ is the
normalized density function, which is 1 for $R_N-\delta < R < R_N + \delta $,
where $R_N$ is the nuclear
radius and $\delta$ is a small number. The single electron matrix elements of $H_{\rm SM}$ are
\begin{equation}
  \langle\psi_{ks_{1/2}}|H_{\rm SM}|\psi_{mp_{1/2}}\rangle=
  3Se\left(-1/3\right)\int\limits_0^\infty \left [P_{ks_{1/2}}(r)
  P_{mp_{1/2}}(r)+ Q_{ks_{1/2}}(r) Q_{mp_{1/2}}\left(r\right)\right] 
  \frac{\rho(R)}{B}RdR .
\end{equation}
where $P_i$ and $Q_i$ are the large and small components of the relativistic 
Dirac-wavefunctions $|\psi_i\rangle $.
The matrix elements of $H_{\rm SM}$ between the states $|\psi_{mp_{1/2}}\rangle$
and $|\psi_{ks_{1/2}}\rangle$ is identical to the above expression.


\section{Coupled-cluster theory for closed-shell atoms}

The starting point of obtaining the coupled-cluster equations is the 
relativistic atomic Hamiltonian in the Dirac-Coulomb approximation
in atomic units ($m_e = 1$, $|e| = 1$ and $\hbar = 1$)
\begin{equation}
   H = \sum_i^N \left [c \mbox{\boldmath{$\alpha_i$}} \cdot 
       \mbox{\boldmath $p_i$} +({\bf\beta_i - 1}) c^2+ 
       V_N(r_i)\right ] + \sum_{i<j}^N \frac{1}{r_{ij}},
  \label{eqdcH}
\end{equation}
where c is velocity of light, $\alpha$ and $\beta$ are the Dirac matrices,
$1/r_{ij}$ is the Coulomb potential energy between two electrons, 
and $V_N(r_i)$ is the nuclear 
potential. In the above Hamiltonian, the rest mass energy is subtracted from 
the total energy eigen values. This is the Hamiltonian of an atomic system 
considering only the electrostatic interactions. The single 
particle equations are obtained by approximating the two-electron term in 
Eq.($\ref{eqdcH}$) by a central field potential $U_{\rm{DF}}(r)$, known as the 
Dirac-Fock potential \cite{lindgren,szabo}, then
\begin{equation}
  H_{\rm{DC}} = \sum_i^N \left [c \mbox{\boldmath{$\alpha_i$}} \cdot \mbox{
                \boldmath $p_i$} + ({\bf\beta_i}-1) c^2+ V_N(r_i) + 
                U_{\rm DF}(r_i) \right] + V_{\rm es}.
\label{eqdc}
\end{equation}
The residual Coulomb interaction
$$
  V_{\rm {es}} = \sum_{i<j}^N \frac{1}{r_{ij}} - \sum_i U_{\rm DF}(r_i).
$$
The non-central (or) correlation effects are included by treating $V_{\rm es}$
as a perturbation. The single electron wavefunctions $\psi_a\rangle $ satisfy 
the Schroedinger equation
\cite{lindgren,szabo}
\begin{eqnarray}
   \left [c \mbox{\boldmath{$\alpha_i$}} \cdot \mbox{\boldmath $p_i$}
    +({\bf\beta_i}-1) c^2+ V_N(r_i) + U_{DF}(r_i) \right]|\psi_a\rangle= 
   \epsilon_a |\psi_a\rangle
\label{dfeqn}
\end{eqnarray}
and are expressed in the relativistic framework in terms of the two-component 
form disucssed in succeeding sections. The Eq.($\ref{dfeqn}$) is obtained 
by variational extremization \cite{lindgren}. The perturbed equations are
obtained by introducing the $P$ and $T$ violating tensor-pseudo tensor 
interaction Hamiltonian in addition to the residual Coulomb interaction.
The exact atomic wavefunction in CC formalism 
$|\Psi\rangle = e^T|\Phi_0\rangle $, where $|\Phi_0\rangle $ is the 
reference state, $T = T_1 + T_2 + \cdots + T_N$ is the cluster 
operator. In our calculations, we use the singles and doubles approximation
$T = T_1 + T_2$, the second quantized form of the operators are
\begin{eqnarray}
  T_1 & = & \sum_{\rm a,p}a_p^\dagger a_a {\bf t}_a^p |\Phi_0\rangle
                 \nonumber \\
  T_2 & = & \sum_{\rm a,p,b,q}\frac{1}{2!}\:a_p^\dagger a_q^\dagger a_b
          a_a{\bf t}_{ab}^{pq}|\Phi_0\rangle .
\end{eqnarray}
The many-body Schroedinger equation of the Dirac-Coulomb Hamiltonian for an 
atomic system, in a state $|\Psi\rangle$ in the CC formalism 
$$ 
  H e^T |\Phi_0\rangle= E e^T |\Phi_0\rangle .
$$
Operating from left side by $e^{-T}$
$$  
  e^{-T} H e^T|\Phi_0\rangle = E|\Phi_0\rangle .
$$
Expressing $H$ in normal ordered form, $H= H_N +E_{DF}$, where 
$E_{DF}= \langle \Phi_0|H|\Phi_0 \rangle - \langle \Phi_0|H_N|\Phi_0\rangle$ 
is the Dirac-Fock energy. Then,
\begin{equation}
  e^{-T}(H_N + E_{DF}) e^T |\Phi_0\rangle= E |\Phi_0\rangle
  \label{normal}
\end{equation}
Projecting Eq.($\ref{normal}$) with singly and doubly excited states
$\langle\Phi^r_a|$ and $\langle\Phi^{rs}_{ab}|$ respectively, 
the single and double 
excitation cluster amplitude equations are \cite{mywork}
\begin{eqnarray}
  \langle\Phi^r_a|\{(\ob{H_Ne^T})_{\rm c}\}|\Phi_0\rangle
  &=& 0    \\
  \langle \Phi^{rs}_{ab}|\{(\ob{H_Ne^T})_{\rm c}\}
  |\Phi_0 \rangle & =& 0 .
\label{genCC}
\end{eqnarray}


\subsection{EDM perturbed coupled-cluster equations}

The $\hedm $ operator is the general EDM operator. In the present paper, $\hedm $ 
denotes $H_{\rm SM}$, which is a single-particle operator.  Consider the 
NSM perturbed Schroedinger equation
\begin{equation}
   {\widetilde H}|{\widetilde \Psi}\rangle= E |{\widetilde\Psi}\rangle
\end{equation}
where ${\widetilde H}=H_{\rm DC}+\lambda H_{\rm EDM}$ and 
$|{\widetilde \Psi}\rangle=e^T|\Phi_0\rangle= e^{T^{(0)}+\lambda T^{(1)}}|\Phi_0\rangle$.
Taking upto one order in $\lambda$ in the exponent,
$$
  {\widetilde H}e^{T^{(0)}}\left(1+\lambda T^{(1)}\right)|\Phi_0\rangle= 
  E e^{T^{(0)}} \left(1+\lambda T^{(1)}\right)|\Phi_0\rangle .
$$
Substituting for ${\widetilde H}$ in the above equation and comparing 
zeroth and first order terms of $\lambda $ on both sides,
\begin{equation}
   \left(H_{\rm DC} e^{T^{(0)}}\right)|\Phi_0\rangle = E e^{T^{(0)}}|\Phi_0\rangle ,
\label{unp}
\end{equation}
and
\begin{equation}
   \left(H_{\rm DC}e^{T^{(0)}}T^{(1)} + H_{\rm EDM}e^{T^{(0)}}\right)|\Phi_0\rangle
   = E e^{T^{(0)}} T^{(1)} |\Phi_0\rangle .
\label{lam1}
\end{equation}
Multiplying Eq.($\ref{unp}$) by $T^{(1)}$ on both sides and substitute 
$H_{\rm DC}=H_N+E_{DF}$ 
\begin{equation}
   T^{(1)} H_Ne^{T^{(0)}}|\Phi_0\rangle = \Delta E_{\rm corr}T^{(1)}e^{T^{(0)}}
   |\Phi_0\rangle .
\label{neweqnunp1}
\end{equation}
and Eq.($\ref{lam1}$) can be written as
\begin{equation}
   \left(H_N e^{T^{(0)}}T^{(1)}+H_{\rm EDM}e^{T^{(0)}}\right)|\Phi_0\rangle
   = \left(\Delta E_{\rm corr} e^{T^{(0)}} T^{(1)}\right)|\Phi_0\rangle
\label{lam2}
\end{equation}
since $T^{(0)}$ and $T^{(1)}$ commute and $\Delta E_{\rm corr} = E - E_{DF}$ is the
correlation energy.
Operate Eq.($\ref{lam2}$) on both sides by $e^{-T^{(0)}}$
\begin{equation}
   \left(\overline H_N T^{(1)} + \overline H_{\rm EDM}\right)|\Phi_0\rangle =
   \Delta E_{\rm corr} T^{(1)}|\Phi_0\rangle
\label{newlam2}
\end{equation}
Operating by $e^{-T^{(0)}}$ on Eq.$\ref{neweqnunp1}$ 
\begin{equation}
   T^{(1)} \overline H_N|\Phi_0\rangle = \Delta E_{\rm corr}T^{(1)}
   |\Phi_0\rangle .
\label{neweqnunp}
\end{equation}
Subtracting Eq.($\ref{newlam2}$) from Eq.($\ref{neweqnunp}$)
\begin{equation}
   \left[{\overline H}_N,T^{(1)} \right] |\Phi_0\rangle = 
   -{\overline H}_{\rm EDM} |\Phi_0\rangle ,
\label{beq}
\end{equation}
where ${\overline O} = e^{-T^{(0)}} \hat O e^{T{(0)}}$ and $\hat O$ is any
operator. The equation for the $\hedm $ perturbed singles and doubles 
cluster amplitudes can be derived from the basic equation, Eq.($\ref{beq}$) 
by projecting on both 
sides of the equation with singly and doubly excited determinantal states.
\begin{eqnarray}
   \langle\Phi_a^r|\left[\overline H_N,T^{(1)}\right]|\Phi_0\rangle & = &
   -\langle\Phi_a^r|\overline H_{\rm EDM}|\Phi_0\rangle
      \nonumber \\
   \langle\Phi_{ab}^{rs}|\left[\overline H_N,T^{(1)}\right]|\Phi_0\rangle & = &
   -\langle\Phi_{ab}^{rs}|\overline H_{\rm EDM}|\Phi_0\rangle 
\end{eqnarray}
which are equivalent to
\begin{eqnarray}
   \langle\Phi_a^r|\left\{\ob{\overline H_NT^{(1)}}\right\}|\Phi_0\rangle & = &
   -\langle\Phi_a^r|\overline H_{\rm EDM}|\Phi_0\rangle ,
      \nonumber  \\
   \langle\Phi_{ab}^{rs}|\left\{{\ob{\overline H_NT^{(1)}}}\right\}
   |\Phi_0\rangle & = & -\langle\Phi_{ab}^{rs}|\overline H_{\rm EDM}|\Phi_0\rangle .
\label{ccedm-eqns}
\end{eqnarray}
Further expanding $T^{(1)}=T_1^{(1)} + T_2^{(1)}$, these equations can 
be cast in the form of a system of linear matrix equations
\begin{eqnarray}
  H_{11} T_1^{(1)} + H_{12} T_2^{(1)} & = & -H_{10},
  \nonumber  \\
  H_{21} T_1^{(1)} + H_{22} T_2^{(1)} & = & -H_{20},
\label{pertmatrix}
\end{eqnarray}
where $H_{11}$, $H_{12}$, $H_{21}$, $H_{22}$ are the sub blocks of the dressed 
$H_N$ matrix elements and $H_{10}$ and $H_{20}$ are the sub blocks of the 
dressed $\hedm $ matrix elements.


\subsection{Calculation of Atomic Electric Dipole moments}

The EDM of the atom in the state $|{\widetilde \Psi} \rangle$ is

\begin{equation}
   D_{\rm atom} = \frac{\langle {\widetilde \Psi}|D|{\widetilde \Psi}
         \rangle }{\langle {\widetilde \Psi}|{\widetilde 
         \Psi} \rangle }
 \label{edmeqn}
\end{equation}
where D is the electric dipole operator.
Substituting the above expression for $|{\widetilde \Psi}\rangle $ 
and keeping terms only of order $\lambda $, we obtain the 
expression for EDM (the symbol of contraction appears using 
$\overline{D}T^{(1)} = \ob{\overline{D}T^{(1)}}+ 
\left\{\overline{D}T^{(1)}\right\}$, where the curly brackets
refer to normal ordering and the expectation value of normal ordered operator 
between the vacuum states is zero)
\begin{equation}
   D_{\rm atom} = \frac{\langle \Phi_0|\left[{\overline D}T^{(1)} + {T^{(1)}}^\dagger
     {\overline D}\right]|\Phi_0\rangle } {\langle \Psi_0|\Psi_0\rangle },
\end{equation}
which can also be written as
\begin{equation}
   D_{\rm atom} = \frac{\langle \Phi_0|[\ob{\overline{D}T^{(1)}} + 
         \ob{{T^{(1)}}^\dagger \overline{ D}}]|\Phi_0\rangle } 
        {\langle \Psi_0|\Psi_0\rangle },
  \label{edmeqn1}
\end{equation}
where $\overline{ D} = {e^{T^{(0)}}}^{\dagger}De^{T^{(0)}}$. Using the fact
that $T^{(1)}$ and $D$ operators are odd and $T^{(0)}$ operator is even under 
parity, the bra and the ket vectors in the above expression must have the same 
parity. To simplify the calculations, we expand $\overline{D}$ as  
\cite{latha-thesis,mywork}
\begin{equation}
   \overline{ D} = \left(1 + {T^{(0)}}^\dagger + 
   \frac{{{T^{(0)}}^\dagger}^2}{2!} + \cdots \right) D e^{T^{(0)}}=
    De^{T^{(0)}} + \sum_{n=1}^{\infty} \frac{1}{n!}
                  \left ({T^{(0)}}^\dagger\right)^n ~ D e^{T^{(0)}}
\label{basiceqn}
\end{equation}
In Eq.($\ref{basiceqn}$), the one-body nature of the electric dipole operator 
$D$ restricts the maximum possible contractions with $T^{(0)}$ to just two. 
Define
$$
  |\Phi_1\rangle = T^{(1)}|\Phi_0\rangle = \left(T_1^{(1)} +
  T_2^{(1)}\right) |\Phi_0\rangle,
$$
then
\begin{equation}
  D_{\rm atom} = \frac{\langle\Phi_0|\overline{ D}|\Phi_1\rangle +
              \langle\Phi_1|\overline{ D}|\Phi_0\rangle}
             {\langle\Psi_0|\Psi_0\rangle}
      =  2\frac{\langle\Phi_1|\overline{ D}|\Phi_0\rangle}
                  {\langle\Psi_0|\Psi_0\rangle}.
\end{equation}
The last step follows as the two terms are the complex conjugates of each other
and give equal contributions. Substituting the expanded form of
$\overline{ D}$
\begin{eqnarray}
 D_{\rm atom} & = & 2\langle\Phi_1|\left[D e^{T^{(0)}} + \sum_{n=1}^{\infty}
           \frac{1}{n!} \left ({T^{(0)}}^\dagger\right )^n De^{T^{(0)}}\right ]
           |\Phi_0\rangle/\langle \Psi_0|\Psi_0\rangle \nonumber  \\
     & = & 2 \left [ \langle \Phi_1|D e^{T^{(0)}}|\Phi_0\rangle +
           \langle \Phi_1|\sum_{n=1}^{\infty} \frac{1}{n!}
           \left ({T^{(0)}}^\dagger\right )^n D e^{T^{(0)}}|\Phi_0\rangle 
           \right ] /\langle \Psi_0|\Psi_0\rangle 
   \label{eqn-final}
\end{eqnarray}
The complexity of the above expression can be mitigated by exploiting the fact
that not all the terms containing the ${T^{(0)}}^\dagger$ operators contribute 
to the infinite summation.  In this scheme, the zeroth order $D_{\rm atom}$ 
consists of terms
without ${T^{(0)}}^\dagger$ operator, first order $D_{\rm atom}$ consists of 
terms having one
order of ${T^{(0)}}^\dagger$, and so on. The unlinked terms of the numerator
cancel with the denominator and only linked terms contribute in the numerator.
At the linear level,
\begin{eqnarray*}
  {\overline D}& = & \left(1 + T^{(0)} \right)^\dagger D \left(1 + 
                        T^{(0)} \right) \\
   & = &  D + DT^{(0)} + {T^{(0)}}^\dagger D \\
   & = &  D + DT_1^{(0)} + 
      DT_2^{(0)} + {T_1^{(0)}}^\dagger D + {T_2^{(0)}}^\dagger D
\end{eqnarray*}
Following are the terms contributing to the EDM expectation value at the 
linear level :
\begin{equation}
  D_{\rm atom} = \langle \Phi_0|{\overline D}T^{(1)}+
    {T^{(1)}}^\dagger{\overline D} |\Phi_0\rangle 
\label{exptval}
\end{equation}
The two terms in Eq.$\ref{exptval}$ are complex conjugates of each other.
Hence
$$
  D_{\rm atom} = 2 \: \langle \Phi_0|{T^{(1)}}^\dagger{\overline D} 
                 |\Phi_0\rangle .
$$
Substituting for $\overline D$ the expression for the atomic EDM
\begin{equation}
   D_{\rm atom} = 2 \langle \Phi_0|\left[{T_1^{(1)}}^\dagger D + 
   {T_1^{(1)}}^\dagger D T_1^{(0)}+{T_1^{(1)}}^\dagger D T_2^{(0)} + 
   {T_2^{(1)}}^\dagger D T_2^{(0)} + {T_2^{(1)}}^\dagger D T_1^{(0)}\right]
   |\Phi_0\rangle
\label{atom_sample}
\end{equation}


\section{Results}

\subsection{Basis}
  The single particle orbitals for all calculations in the subsequent sections 
were generated using the Gaussian basis \cite{kim} set expansion whose salient features
are presented in this section. In the central field approximation, the solution
of the Dirac equation in terms of the four component spinors is given by
\begin{equation*}
  \psi_{n\kappa m}(r, \theta, \phi) = r^{-1} \left(
                      \begin{array}{c}
                             P_{n\kappa}(r) \:\chi_{\kappa\; m}(\theta, \phi) \\
                                               \\
                            iQ_{n\kappa}(r)\: \chi_{-\kappa\; m}(\theta, \phi)
                      \end{array}
                      \right)
\end{equation*}
where $P_{n\kappa}(r)$ and $Q_{n\kappa}(r)$ are the large and the small 
components of the radial wavefunctions expanded in terms of the basis 
sets\cite{clementi},
\begin{eqnarray*}
   P_{n\kappa}(r) &  =  &  \sum_p C_{\kappa p}^L \: g_{\kappa p}^L(r) \\
   Q_{n\kappa}(r) &  =  &  \sum_p C_{\kappa p}^S \: g_{\kappa p}^S(r)
\end{eqnarray*}
where the summation over the index $p$ runs over the number basis functions $N$,
$g_{\kappa p}^L(r)$ and $g_{\kappa p}^S(r)$ correspond to the large and small
components and $C_{\kappa p}^L$ and $C_{\kappa p}^S$ are their expansion 
coefficients for each value of $\kappa $. The functions $g_{\kappa p}^L(r)$ 
are chosen to be the two parameter Gaussian Type Orbitals (GTOs) \cite{clementi} and 
large and small components are related by the condition of kinetic balance
\cite{stanton,grantkb}.
The calculations presented in this paper correspond to the Even Tempered (ET) basis
set. For the nuclear density we use the two parameter Fermi distribution
$$
     \rho_N(r) = \frac{\rho_0}{1 + e^{(r - c)/a}}
$$
where $r$ is the radial coordinate, $a= 2.3/4 {\rm ln}3 $ is the skin thickness
parameter and $c=\sqrt{\left(5r^2_{\rm rms}/3 - 7a^2\pi^2/3 \right)}$, the half-charge
radius \cite{parpia-nuc}. In the expression of $c$, the $r_{\rm rms}$ is the nuclear 
mean-square radius. The quantity $\rho_0$ is determined by normalising $\rho_N(r)$ over a
spherical volume.  The radial grid has the form,
$r_k=r_0\left(e^{(k-1)h} - 1\right), k=1,.....,n_p$, $n_p$ is the total number
of grid points.

\begin{table}[hb]
\caption{\label{gauss-table}No. of basis functions used to generate the even
tempered Dirac-Fock orbitals and the corresponding value of $\alpha_{0}$
and $\beta$ used. The total number of active orbitals are shown in the brackets
for {\textit Active holes}.}
\begin{center}\begin{tabular}{llllllllll}
\br 
&
$s_{1/2}$&
$p_{1/2}$&
$p_{3/2}$&
$d_{3/2}$&
$d_{5/2}$&
$f_{5/2}$&
$f_{7/2}$&
$g_{7/2}$&
$g_{9/2}$\tabularnewline
\mr
Number of basis&
31&  32& 32& 20& 20& 20& 20& 10&  10
\tabularnewline
$\alpha_{0}(\times10^{-5})$&
725& 715& 715& 700& 700& 695& 695& 655& 655
\tabularnewline
$\beta$&
2.725& 2.715& 2.715& 2.700& 2.700& 2.695& 2.695& 2.655& 2.655
\tabularnewline
Active holes (36) &
2& 2& 2& 2& 2& 1& 1& 1& 1
\tabularnewline
Active holes (39) &
3& 3& 3& 2& 2& 1& 1& 1& 1
\tabularnewline
Active holes (43) &
3& 3& 3& 3& 3& 2& 2& 1& 1
\tabularnewline
Active holes (45) &
3& 3& 3& 3& 3& 3& 3& 1& 1
\tabularnewline
Active holes (51) &
5& 5& 5& 3& 3& 3& 3& 1& 1
\tabularnewline
Active holes (57) &
7& 7& 7& 3& 3& 3& 3& 1& 1
\tabularnewline
Active particles&
6& 4& 4& 3& 3& 1& 1& 0& 0
\tabularnewline
\br
& & & & & & & & &
\tabularnewline
\end{tabular}
\end{center}
\end{table}
For the present calculation, the details of the basis set used in
generating the orbitals is shown in Table. $\rm \ref{gauss-table}$. Using the above basis, the single
particle orbitals have been generated. Next, an existing closed-shell 
coupled-cluster code developed inhouse has been used to generate the 
unperturbed cluster amplitudes \cite{rajat}. The code for calculating 
the perturbed cluster amplitudes, and subsequently the atomic EDM was 
developed by our group \cite{latha-thesis,mywork}. 

\subsection{Results for Hg EDM induced by the ${\hat P}$ and ${\hat T}$
            violating T-PT interaction}
\label{sample}

Contributions from each of the terms in Eq.($\ref{atom_sample}$) are shown
in Table.$\rm \ref{lcc-result}$ for the basis with 57 active orbitals. 
\begin{table}
\caption{\label{lcc-result}Individual contributions}
\lineup
\begin{center}\begin{tabular}{ll}
\br
&
Contributions in units \tabularnewline 
& of e $a_0 \: G_F \: C_T $
\tabularnewline
\mr
${T_1^{(1)}}^\dagger D$ &
\-47.830
\tabularnewline
${T_1^{(1)}}^\dagger D T_1^{(0)}$&
\00.082
\tabularnewline
${T_1^{(1)}}^\dagger D T_2^{(0)}$&
14.320
\tabularnewline
${T_2^{(1)}}^\dagger D T_2^{(0)}$&
\0\-0.386
\tabularnewline
${T_2^{(1)}}^\dagger D T_1^{(0)}$&
\0\-0.059
\tabularnewline \mr
Total &
 \-33.874
\tabularnewline
\br
&
\tabularnewline
\end{tabular}
\end{center}
\end{table}
The final result can be expressed in units of e-m
$$
   D_{\rm Hg} = - 1.125 \times 10^{-22} C_T {\mbox e m} \sigma_N.
$$
The Dirac-Fock contribution is $-2.45 \times 10^{-22} {\rm e m} C_T \sigma_N$.
It can be noticed from the Table $\rm \ref{lcc-result}$  that the largest 
contribution is from ${T_1^{(1)}}^\dagger D$. This can be explained
from the fact that it includes the Dirac-Fock effect, which is the most dominant 
contribution. The trend shown by the individual contributions in 
Table.$\rm \ref{lcc-result}$ is related to the fact that the $T_1^{(0)}$ cluster
amplitudes are smaller in magnitude compared to the $T_2^{(0)}$ cluster 
amplitudes. In addition, the $T_1^{(1)}$ amplitudes are larger in magnitude compared 
to the $T_2^{(1)}$ amplitudes again due to the presence of the
Dirac-Fock contribution in $T_1^{(1)}$. For example, from the above table, we
see that the contribution of ${T_1^{(1)}}^\dagger D T_2^{(0)}$ is larger
than that of ${T_1^{(1)}}^\dagger D T_1^{(0)}$. Similarly, the contribution of 
${T_2^{(1)}}^\dagger D T_2^{(0)}$ is greater than that of 
${T_2^{(1)}}^\dagger D T_1^{(0)}$. These arguments are equally valid for the 
atomic EDM induced by the nuclear Schiff moment shown in the next section, 
which follows the same trend as above. It is also interesting to note that 
in both cases contribution of the term ${T_1^{(1)}}^\dagger D$ is approximately
3 times larger in magnitude than that of the term ${T_1^{(1)}}^\dagger 
D T_2^{(0)}$, which is the second largest contribution.

\subsection{Results for Hg EDM induced by the ${\hat P}$ and ${\hat T}$
            violating Nuclear Schiff moment}
\label{sampleNSM}

The $^{199}{\rm Hg}$ atomic EDM induced by the nuclear Schiff moment is calculated 
with the same basis given in the previous section. The methods of 
generating the perturbed and the unperturbed cluster amplitudes are the 
same as described earlier. The Dirac-Fock contribution is 
$D_{\rm Hg}=-0.546\times 10^5 e\;a_0\: S =-0.390 \times 10^{-17}{\rm e cm}
S(e fm^3)^{\scriptscriptstyle -1} $.
Contributions from each of the terms in Eq.$\ref{atom_sample}$ is shown in 
Table.$\rm \ref{lcc-result-nsm}$. Again, the leading contribution is from
${T_1^{(1)}}^\dagger D$ and it can also be seen that the contribution 
of ${T_1^{(1)}}^\dagger D$ is approximately 3 times in magnitude of  
the contribution of ${T_1^{(1)}}^\dagger D T_2^{(0)}$, which is also true for the 
results of the 
T-PT induced EDM. Our result is not in agreement with Dzuba et.al \cite{dzuba}.
They have used CI + MBPT method for the generation of the orbitals.
We have compared the Schiff moment interaction and the electric dipole matrix 
elements of the $6s_{1/2}$ and the core $p_{1/2}$ orbitals for $^{199}{\rm Hg}$ with 
the results obtained by the authors of Dzuba et al. at the Dirac-Fock level 
and found that the agreement was very good. This suggests that the discrepancy 
in our results could be majorly due to the different choices of the virtual 
orbitals in the two calculations. 

\begin{table}
\caption{\label{lcc-result-nsm}Individual contributions}
\lineup 
\begin{center}\begin{tabular}{ll}
\br
&
Contributions in atomic \tabularnewline 
& units of $S$ e $\times a_0$
\tabularnewline
\mr
${T_1^{(1)}}^\dagger D$ &
\0\-0.177 $\times 10^5 $
\tabularnewline
${T_1^{(1)}}^\dagger D T_1^{(0)}$&
\00.030 $\times 10^3$
\tabularnewline
${T_1^{(1)}}^\dagger D T_2^{(0)}$&
\00.525 $\times 10^4$
\tabularnewline
${T_2^{(1)}}^\dagger D T_2^{(0)}$&
\-14.258 $\times 10^1$
\tabularnewline
${T_2^{(1)}}^\dagger D T_1^{(0)}$&
\00.252 $\times 10^{-4}$
\tabularnewline \mr
Total &
\0\-0.126 $\times 10^5 $
\tabularnewline
\br
&
\tabularnewline
\end{tabular}
\end{center}
\end{table}
The final result is in the units where $S$ is expressed in $\textrm e fm^3$
$$
   D_{\rm Hg} = -0.126 \times 10^5 \times \frac{2 \times 10^{-23}}{0.529^2} \times
               \frac{S}{\mbox efm^3} {\mbox e cm}
$$
$$
   D_{\rm Hg} = - 0.0901 \times 10^{-17} {\rm e cm} \frac{S}{\mbox e fm^3}
$$
\subsection{Summary of the EDM results of $^{199}{\rm Hg}$}
The interaction Hamiltonian for the T-PT and
the NSM are both dependent on the nuclear density $\rho_N(r)$ and hence
their matrix elements are sensitive to the $s_{1/2}$ and the $p_{1/2}$
orbitals, which have a non-zero probability density inside the nuclear
radius. In addition, the NSM interaction Hamiltonian is proportional to the
electron coordinate ${\vec R}$. In addition, the matrix elements of the atomic EDM
contain the electric dipole operator ${\vec D}$. Hence the atomic EDMs
arising from these interactions require the single particle orbitals to be
very accurate at all radial ranges. The trend followed by the
T-PT and NSM induced EDM shows that the atomic EDM is very
sensitive to the inclusion of $s_{1/2}$ and the $p_{1/2}$ virtuals.
On the other hand, there is not much variation in the polarizability as
it is more sensitive to orbitals of higher orbital angular
momenta. We have also performed linear CCEDM calculations without
including any $T^{(0)}$ cluster amplitudes. This in other words, amounts
to the linear CCEDM calculation with the cluster amplitudes generated with
the approximations $\overline H_N \approx H_N$ and $\overline H_{\rm EDM}
\approx H_{\rm EDM}$ in Eq.($\ref{ccedm-eqns}$). Hence all the correlation 
effects from $T^{(0)}$ amplitudes are omitted.
The results of this calculation for a basis of 39 active orbitals are
discussed. The Dirac-Fock contribution for this basis is $D_{\rm Hg}
= -3.17 \times 10^{-22} {\rm e m}$. The total contribution is 
$D_{\rm Hg} = -5.77 \times 10^{-22} {\rm e m}$, whereas the CPHF 
calculation gives $D_{\rm Hg} = -4.64 \times 10^{-22} {\rm e m}$. Similar 
comparison can be made for a basis of 57 active orbitals, for which the 
Dirac-Fock contribution is $D_{\rm Hg} = -2.45 \times 10^{-22} {\rm e m}$, the 
bare-Coulomb calculation without $T^{(0)}$ amplitudes gives 
$D_{\rm Hg} = -6.60 \times 10^{-22} {\rm e m}$, the CPHF calculation gives 
$D_{\rm Hg} = -5.61 \times 10^{-22} {\rm e m}$. This comparison helps in 
understanding the interplay between the various many-body effects :  the 
the bare-Coulomb which contains all kind of correlation effects to one order 
in Coulomb interaction;the CPHF theory which contains the Coulomb effects only 
of the two particle-two hole kind; the linear CCEDM which subsumes the effects 
of the bare Coulomb, the CPHF and more.

\begin{table}
\caption{Summary of our results.}
\lineup
\begin{center}\begin{tabular}{lllll}
\br
&
Basis size&
In units of &
In units of 
\tabularnewline
&
&
$10^{-22}C_T {\rm e m}$ &
$10^{-17}$ {\rm e cm} $\frac{S}{\rm e fm^3}$
\tabularnewline
\mr
& 36 & \-2.70 &  1.83 
\tabularnewline
& 39 & \-2.97 & \-0.23 
\tabularnewline
& 42 & \-2.21 & \-0.15 
\tabularnewline
& 45 & \-1.41 & \-0.11 
\tabularnewline
& 51 & \-1.40 & \-0.11 
\tabularnewline
& 57 & \-1.13 & \-0.09 
\tabularnewline
\br
& & & &
\tabularnewline
\end{tabular}
\end{center}
\end{table}

Limits on the P and T violating parameters at the elementary particle level can be
derived from the tensor-pseudo tensor coupling constant ($C_T$) and the
nuclear Schiff moment ($S$)\cite{latha-thesis,mywork}. The results of various P and T
violating parameters are shown in Table.${\rm \ref{cpvp}}$. A detailed discussion on the
derivation of these parameters has been presented in \cite{latha-thesis}. Using these
important parameters, the observable EDMs can be connected to the P and T violating 
coupling constants predicted by the underlying elementary particle theories. The bounds
on these coupling constants can further be used to constrain various models of elementary 
particle physics.

\begin{table}
\caption{P and T violating parameters}
\label{cpvp}
\lineup
\begin{center}\begin{tabular}{l l}
\br
P and T violating parameter at the  &   Limits obtained from \tabularnewline 
elementary particle level      &   linear CCEDM \tabularnewline 
\mr
$\eta_{np}$     &  $ < -1.6 \times 10^{-2}$ 
              \tabularnewline
$\overline g_{\pi NN}$ &   $ < 1.7 \times 10^{-10}$
              \tabularnewline
$\theta_{\rm QCD}$  & $ < 6.3  \times 10^{-9}$
              \tabularnewline
e$\left(\bar d_u - \bar d_d\right)$  &  $ < 0.86 
      \times 10^{-24}$ e cm \tabularnewline \br
\end{tabular}
\end{center}
\end{table}

\end{document}